\documentclass[pdflatex,sn-vancouver-num]{sn-jnl}

\usepackage{geometry}
\usepackage{graphicx}%
\usepackage{multirow}%
\usepackage{amsmath,amssymb,amsfonts}%
\usepackage{amsthm}%
\usepackage{mathrsfs}%
\usepackage{xcolor}%
\usepackage{textcomp}%
\usepackage{manyfoot}%
\usepackage{booktabs}%
\usepackage{pifont}

\usepackage[utf8]{inputenc}
\usepackage[english]{babel}
\usepackage[strict]{csquotes}
\usepackage{tcolorbox}
\usepackage{soul}

\usepackage{tabularx}
\usepackage{longtable}

\usepackage{hyperref}
\hypersetup{bookmarksdepth=2} 
\usepackage{makecell}
\usepackage{caption}
\usepackage{array}
\usepackage{colortbl}
\usepackage{float}
\usepackage{pifont}
\usepackage{xurl}
\makeatletter
\g@addto@macro{\UrlBreaks}{%
  \do\a\do\b\do\c\do\d\do\e\do\f\do\g\do\h\do\i\do\j\do\k\do\l\do\m
  \do\n\do\o\do\p\do\q\do\r\do\s\do\t\do\u\do\v\do\w\do\x\do\y\do\z
  \do\A\do\B\do\C\do\D\do\E\do\F\do\G\do\H\do\I\do\J\do\K\do\L\do\M
  \do\N\do\O\do\P\do\Q\do\R\do\S\do\T\do\U\do\V\do\W\do\X\do\Y\do\Z
  \do\0\do\1\do\2\do\3\do\4\do\5\do\6\do\7\do\8\do\9}
\makeatother

\usepackage{xcolor}
\definecolor{mySkyBlue}{RGB}{135,206,235}

\newcommand{\cmark}{\ding{51}}

\newcommand{\ours}{\texttt{MedVLMBench}}

\newcommand{\citesupp}[1]{\cite{#1}}

\theoremstyle{thmstyleone}%
\theoremstyle{thmstyletwo}%

\theoremstyle{thmstylethree}%

\begin{document}

\title[Article Title]{Can Generalist Vision Language Models (VLMs) Rival Specialist Medical VLMs? Benchmarking and Strategic Insights}


\author[1]{\fnm{Yuan} \sur{Zhong}}\equalcont{These authors contributed equally to this work.}

\author[2,3]{\fnm{Ruinan} \sur{Jin}}
\equalcont{These authors contributed equally to this work.}

\author*[1]{\fnm{Qi} \sur{Dou}}\email{qidou@cuhk.edu.hk}

\author*[2,3]{\fnm{Xiaoxiao} \sur{Li}}\email{xiaoxiao.li@ece.ubc.ca}

\affil*[1]{\orgname{The Chinese University of Hong Kong}}

\affil[2]{\orgname{The University of British Columbia}}

\affil[3]{\orgname{Vector Institute}}


\abstract{\textbf{Objective}\quad Medical vision-language models (VLMs) are commonly pretrained on large multimodal medical corpora and assumed to outperform generalist VLMs in clinical image understanding. However, specialist development requires substantial curated data, computing resources and engineering expertise. We aimed to determine when pretrained specialist medical VLMs offer practical advantages over generalist models across deployment-relevant medical imaging settings.

\textbf{Methods and analysis}\quad We developed MedVLMBench to compare generalist VLMs with specialist medical counterparts from related model families. The benchmark includes contrastive VLMs for disease diagnosis and generative multimodal large language models for visual question answering. We evaluated 18 VLMs across 12 public dataset components spanning radiology, pathology, dermatology and ophthalmology. Models were assessed under off-the-shelf use, efficient local adaptation, cross-domain transfer to modalities outside specialists’ reported pretraining scope, and agentic reasoning using two training-free multi-agent frameworks. Linear mixed-effects models estimated performance differences across datasets and model families.

\textbf{Results}\quad Specialist medical VLMs showed conditional off-the-shelf advantages, strongest when target modalities matched their reported pretraining scope, but these gains were inconsistent across datasets and model families. Lightly adapted generalist VLMs were generally competitive and often matched or exceeded the strongest off-the-shelf specialist baselines. In cross-domain evaluations, adapted generalists frequently matched or outperformed adapted specialists on unfamiliar or heterogeneous modalities. Under agentic frameworks, specialist and generalist models did not differ significantly in the benefit obtained. Gains were limited and heterogeneous, varying by model family, architecture and pretraining; current specialist models, typically smaller and not optimised for reasoning, benefited little from agentic scaffolding.

\textbf{Conclusion}\quad 
Specialist medical VLMs should not be assumed to be universally superior. Their advantage is most apparent in off-the-shelf, modality-matched deployment, but is largely offset by lightweight adaptation of generalist models and does not consistently extend to cross-domain or agentic settings. Model selection should therefore reflect the intended deployment pathway rather than specialist status alone. Strong generalist VLMs may offer a credible and more scalable alternative, particularly for cross-domain and resource-constrained applications. Progress in clinical agentic AI may depend more on larger reasoning-oriented models and purpose-built medical agent frameworks than on domain-specialised backbones. 
}

\keywords{Vision-Language Models; Multi-modal Agent; Vision Question-Answering; Benchmarking}



\maketitle

\begin{tcolorbox}[colframe=black, colback=gray!5, sharp corners, title={Key messages}]
\textbf{What is already known on this topic}
\begin{itemize}
    \item Specialist medical VLMs are often assumed to outperform generalist models in clinical imaging tasks, but previous studies have provided limited evidence on matched generalist–specialist comparisons, lightweight adaptation, cross-domain transfer, and agentic reasoning.
\end{itemize}
\textbf{What this study adds}
\begin{itemize}
    \item We show that specialist advantages are conditional: they are most evident in off-the-shelf, modality-matched settings, while lightly adapted generalist VLMs often match or exceed specialist baselines and may transfer better to unfamiliar medical imaging domains. 
    \item Under agentic frameworks, specialist and generalist models did not differ significantly, and gains were limited, particularly for current specialist models that are typically small in scale and not optimized for reasoning.
\end{itemize}
\textbf{How this study might affect research, practice or policy}
\begin{itemize}
    \item The study systematically characterizes the differences between generalist VLMs and specialist VLMs extensively pretrained on medical data, and provides practical guidance on model selection for clinical deployment across multiple scenarios from the perspectives of cost and efficiency.
    \item The findings support evaluating medical VLMs by deployment pathway rather than by specialist status alone, with transparent reporting of pretraining scope, possible benchmark exposure, local adaptation requirements and clinically relevant validation before adoption.
\end{itemize}
\end{tcolorbox}

\section{Introduction}
\label{intro}

The rapid evolution of artificial intelligence (AI) in healthcare has accelerated the development of foundational vision–language models (VLMs) for medical imaging: large-scale systems trained for multimodal understanding and generation~\citep{lu2025integrating,zhang2024generalist,moor2023foundation,amugongo2025retrieval,Yime000014}. These models promise a generalist capability across tasks ranging from clinical image-text processing to diagnostic image interpretation by leveraging extensive unsupervised pretraining over large and diverse datasets. Recent advances have produced \textbf{specialist medical VLMs}~\citep{zhang2023biomedclip,xu2025lingshu,sellergren2025medgemma,li2023llava,shi2026medxiaohe,Vishwanathe000068}\footnote{We define specialist medical VLMs as the VLMs pretrained or continue-pretrained primarily on medical corpora.}, adopting extensive pretraining on multimodal medical corpora spanning radiology, pathology, ophthalmology, etc. Despite their promise, such specialist pipelines demand substantial computational resources and curated healthcare data at scale, rendering them expensive and technically challenging to develop~\citep{bommasani2021opportunities}. 
This cost is especially consequential in settings where the resources required to build and maintain specialist models may limit scalability, reproducibility and equitable access to clinical AI tools~\citep{rutunda2026large}. For example, a community hospital or health system in a low-resource country may lack the curated multimodal datasets, compute infrastructure, and specialist engineering support needed to develop a bespoke medical VLM, making it harder to reproduce results locally or to extend clinical AI tools across institutions.


In many real-world clinical applications, AI models are expected to excel at specific tasks, particularly within the data distributions in which they are deployed~\citep{castro2020causality,moor2023foundation}. The reliance on broad generalist models may not always align with the needs of high-stakes, domain-specific applications in medical imaging~\citep{hager2024evaluation,wang2025perspective}. While the prevailing assumption is that models pretrained on medical corpora yield superior generalization~\citep{chia2024foundation,campo2026publicly,sali2026foundation,chen2026visual}, this raises a more precise deployment question: when does an available specialist VLM provide a practical advantage over a domain-agnostic VLM that can be adapted with lightweight fine-tuning? Furthermore, claims of strong transfer in specialist medical VLMs remain difficult to verify, as evaluation outside the specialist's pretraining modalities has been limited~\citep{nan2025beyond}. This leaves open whether observed transfer patterns are consistent with a trade-off between domain depth and transfer breadth.


Recent evidence from clinical language modelling has challenged the assumption that specialization necessarily confers an advantage in medicine. Vishwanath et al.~\cite{vishwanath2026general} found that frontier general-purpose LLMs outperformed specialized clinical AI tools across medical knowledge, clinician-alignment and real clinical query benchmarks, highlighting the need for independent evaluation of specialist systems before clinical deployment. Whether an analogous pattern holds for diverse medical imaging VLMs, however, remains unexplored.

In contrast, \textbf{generalist VLMs}~\citep{radford2021learning,liu2023visual,team2025gemma,bai2025qwen2}, though not originally developed for medicine, are widely available, pretrained on massive general-domain data, and exhibit strong transferable priors. Crucially, these models can often be adapted to target clinical tasks through lightweight approaches (e.g., parameter-efficient fine-tuning) at substantially lower cost~\citep{wang2025perspective}. This motivates broader evaluations beyond off-the-shelf performance. At the same time, recent work has investigated multi-agent systems for medical question answering and decision-support research~\citep{kim2024mdagents,feng2025ucagents}, in which VLMs assume independent roles and engage in structured multi-round interactions, including debate, critique, and consensus-building. This also motivates an assessment of whether costly specialist medical VLMs provide measurable advantages within advanced agentic frameworks.
Together, these developments raise a \textbf{central research question}: \emph{Under what conditions do available pretrained specialist medical VLMs offer practical performance advantages over efficiently adapted generalist VLMs?}

To provide a principled answer to this central question, this paper operationalizes it through four research questions (RQs):
\begin{tcolorbox}[colframe=black, colback=gray!10, sharp corners]
\textbf{RQ1}: For solving specific medical imaging tasks, what is the gap between off-the-shelf specialist medical and generalist VLMs? \\
\textbf{RQ2}: Can a generalist VLM, after simple fine-tuning, achieve comparable or superior performance?
\\
\textbf{RQ3}: For tasks involving unfamiliar imaging modalities, does specialization enhance robustness or limit transfer? \\
\textbf{RQ4}: How do reasoning multi-agent frameworks affect the gap between generalist and medical VLMs?
\end{tcolorbox}

In this paper, we introduce \ours, a benchmark designed to address these gaps. \ours{} evaluates VLMs both in their off-the-shelf (OTS) form and after lightweight adaptation. It supports structured benchmark comparisons between generalist VLMs and specialist medical counterparts within the \emph{same model family}. The benchmark primarily covers two model settings—contrastive VLMs and generative multimodal large language models—and includes both generalist representatives and specialist medical variants. To probe cross-domain transfer, \ours{} explicitly evaluates both tasks whose modality is well-represented in specialist pretraining corpora and tasks involving modalities outside that distribution, thereby testing whether observed performance patterns are consistent with specialist pretraining improving domain depth at the expense of transfer breadth. The benchmark further includes an agentic evaluation that compares generalist and specialist medical VLM backbones within representative training-free multi-agent reasoning frameworks.

Using this framework, we systematically evaluated 18 contrastive and generative VLMs across 12 public dataset components spanning radiology, pathology, dermatology, and ophthalmology, including 96 diagnostic evaluations, 104 standard closed-form VQA evaluations, and 92 agentic VQA evaluations (Figure~\ref{fig:overview}). We find that the off-the-shelf advantage of specialist medical VLMs is task-specific and inconsistent across families. It appears on selected datasets within the specialist's reported training domain, but is not reliably superior at the family level. Under the evaluated lightweight adaptation protocols, generalist VLMs frequently match or surpass off-the-shelf specialist baselines on specialized-domain tasks, and adapted generalists are competitive with or superior to adapted specialists on several cross-domain tasks. 
Under training-free agentic reasoning frameworks, generalist and specialist medical models showed no significant difference in benefit. Gains were limited and heterogeneous, since the current specialist medical models are typically small in scale and not optimized for reasoning, they benefited little from agentic scaffolding, and the magnitude of improvement varied with model family, architecture and pretraining regime.

\section{Methods}
\label{method}

\subsection{Overview and evaluation protocols}
\begin{figure}[t!]
    \centering
    \includegraphics[width=\linewidth]{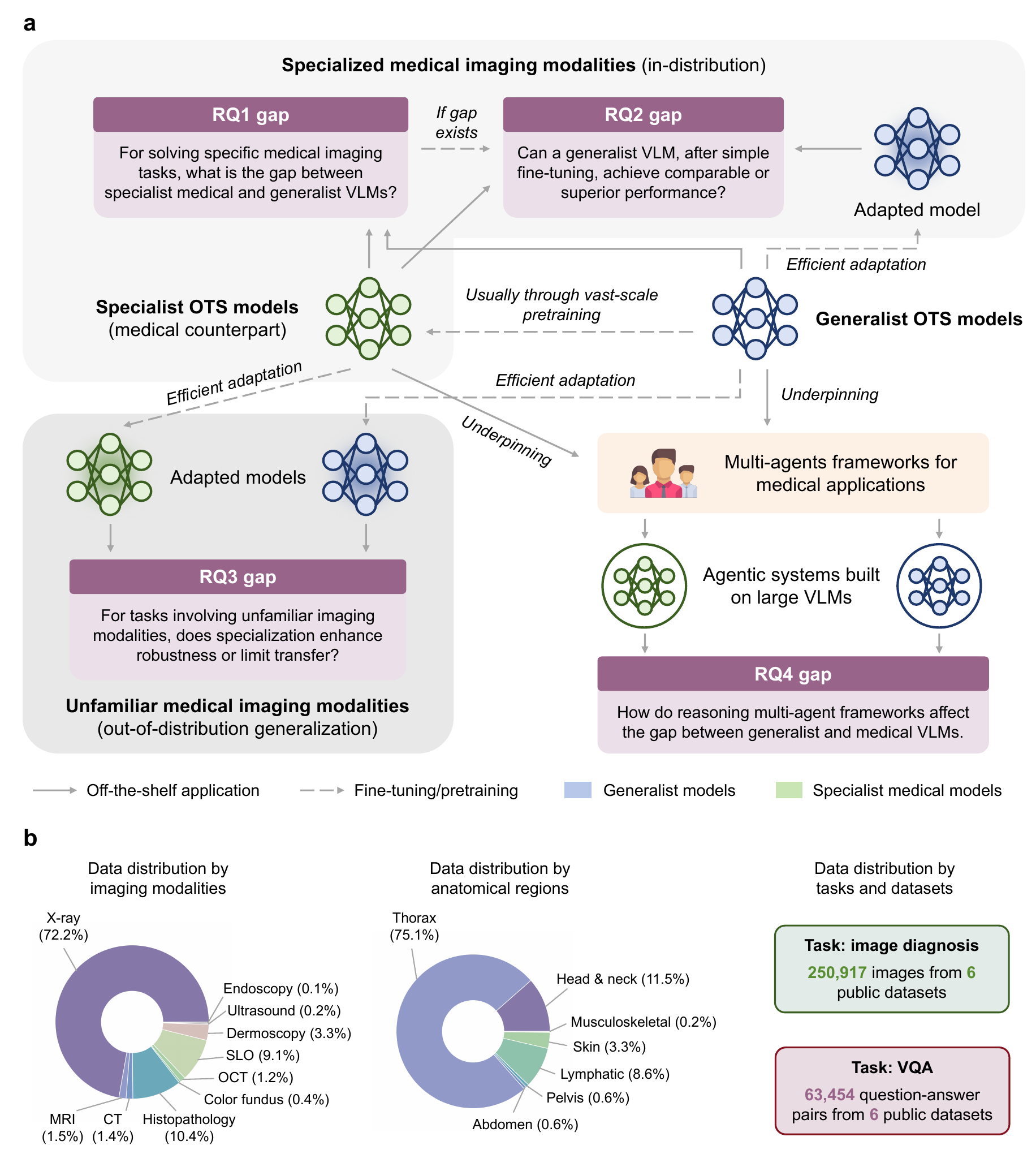}
    \caption{\textbf{Overview of the study.} \textbf{a.} \ours{} investigates when available specialist VLMs provide practical advantages over efficiently adapted generalist VLMs, structured around four RQs. RQ1 assesses how consistent and reliable the OTS specialist advantage is across tasks and model families. RQ2 asks whether lightweight fine-tuning of a generalist model can offset the OTS specialist baseline on specialized-domain tasks in a deployment-oriented comparison. RQ3 examines whether cross-domain transfer patterns are consistent with specialist pretraining limiting representational breadth, by testing transfer to modalities outside the specialist's pretraining distribution. RQ4 evaluates specialist and generalist backbones within representative training-free multi-agent reasoning frameworks. \textbf{b.} Distribution of the benchmark data by imaging modality, anatomical region, and task, covering image diagnosis and VQA across diverse medical datasets.}
    \label{fig:overview}
\end{figure}

This study systematically compares generalist and specialist medical VLMs across a broad range of tasks, data types, and model families, as well as clinically relevant deployment scenarios, including generalization to unseen or rare imaging modalities and agentic reasoning. As shown in Fig.~\ref{fig:overview}, we focus on comparisons between generalist models and their medical counterparts within the same model family and under matched architectural and pretraining conditions, so that differences in performance could be attributed more directly to medical specialization rather than to model choice alone. In parallel, we also assessed whether these findings held in global cross-family comparisons.



Across diverse medical imaging datasets and modalities, we define the \textbf{specialized-domain} setting of a specialist medical VLM as one in which the primary imaging modality is represented within the model's reported medical pretraining scope. By contrast, \textbf{cross-domain} evaluation refers to generalization to unseen or clinically rare imaging modalities. For medical VLMs pretrained on broad medical imaging corpora, such as Lingshu~\cite{xu2025lingshu} and MedGemma~\cite{sellergren2025medgemma}, common modalities, including CT and X-ray, are considered specialized-domain settings because they are consistently covered by the reported pretraining data. Rare modalities, such as SLO fundus imaging, are treated as cross-domain because they are largely absent from reported training cohorts or remain challenging for current VLMs~\cite{qin2025lmod,luo2024harvard}. Datasets comprising heterogeneous and uncommon modalities, such as OmniMedVQA~\cite{hu2024omnimedvqa}, are likewise categorized as cross-domain. For generalist VLMs, all medical imaging datasets are considered cross-domain, given their partially disclosed training recipes and performance gap relative to contemporary medical VLMs on medical tasks~\cite{li2023llava,zhang2023biomedclip,sellergren2025medgemma}.

In the study, four comparison groups were conducted in alignment with the research questions proposed above. 
RQ1 assesses how consistent and reliable the OTS specialist advantage is across zero-shot tasks and model families. RQ2 compares an adapted generalist VLM with an OTS specialist VLM on specialized-domain tasks, reflecting a practical deployment choice between lightweight local adaptation and reliance on specialist medical pretraining.
RQ3 compares the cross-domain transfer of adapted generalist and adapted specialist VLMs to modalities outside the specialist model's pretraining distribution. 
RQ4 evaluates whether training-free multi-agent frameworks alter the comparative performance of generalist and specialist VLMs. 

\subsection{Tasks and datasets}
\label{sec:datasets}

The evaluation spans two task categories, disease diagnosis and VQA, across 12 public datasets in radiology, pathology, dermatology, and ophthalmology. Disease diagnosis is evaluated using the area under the receiver operating characteristic curve (AUROC), and VQA is evaluated under two inference settings: standard prompting and agentic prompting, where pretrained VLMs are augmented with representative multi-agent reasoning frameworks~\cite{kim2024mdagents,feng2025ucagents} to support question answering and decision-making. To ensure a standardized protocol across datasets and model families, we primarily use yes/no and multiple-choice questions, with accuracy as the evaluation metric. Task and dataset details are provided in Table.~\ref{tab:diagnosis-data} and~\ref{tab:vqa-data} in the online supplemental material.
The FairVLMed~\citep{luo2024fairclip} dataset contains both disease diagnosis labels and textual reports. We therefore constructed question-answer pairs from the reports to enable VQA evaluation. Specifically, GPT-4~\cite{openai2025gpt41} was used to generate one closed-form and one open-form question from each clinical report, followed by human validation for answerability, label consistency and clinical plausibility. 
The detailed prompt for VQA construction is described in Fig.~\ref{fig:vqa_prompt} in the online supplemental material.


For both diagnosis and VQA, model selection, lightweight adaptation and final evaluation used the official dataset splits provided by the original benchmark or dataset release. The official test split was held out from fine-tuning and hyperparameter selection and used only for final evaluation. This reduces benchmark-level contamination between adaptation and evaluation data, but it cannot rule out exposure during foundation-model pretraining. We reviewed public model documentation for reported overlap with downstream benchmarks where possible. 
More broadly, large or partially undisclosed pretraining corpora may contain unreported medical samples or near-duplicates. We therefore avoid using single-dataset advantages as definitive evidence of specialization, especially when exposure is known or cannot be excluded.



\subsection{Models and fine-tuning strategies}
We included both contrastive VLMs and generative VLMs (i.e., multimodal large language models, MLLMs), spanning the CLIP~\cite{radford2021learning}, SigLIP~\cite{zhai2023sigmoid}, LLaVA~\cite{li2023llava}, Qwen~\cite{bai2025qwen2} and Gemma~\cite{team2025gemma} families, together with leading medical counterparts within each family, such as BiomedCLIP~\cite{zhang2023biomedclip}, MedSigLIP~\cite{sellergren2025medgemma}, LLaVA-Med~\cite{li2023llava}, Lingshu~\cite{xu2025lingshu} and MedGemma~\cite{sellergren2025medgemma}. These medical counterparts share the same architectures as their generalist counterparts and were obtained through large-scale retraining or continued pretraining on medical multimodal data, enabling their use across a broad range of downstream clinical tasks. Within each model family, we ensured that the included medical and generalist counterparts were of comparable model size. In addition, we included several strong standalone models, including frontier reasoning LLMs such as o3~\cite{openai2025o3o4mini} and Gemini 2.5 Pro~\cite{comanici2025gemini}. 
Detailed information on included models is provided in Table.~\ref{tab:vlm_family_categorization} and \ref{tab:gen_vlm_summary_part2} in the online supplemental material.


\vspace{2mm}
\noindent\textbf{Contrastive VLMs.} Contrastive VLMs map images and text into a common embedding space using contrastive learning to enable zero-shot diagnosis. Our evaluation is structured around two core architectural families. First, the CLIP family is examined, comparing the foundational web-trained CLIP~\cite{radford2021learning} and BLIP2~\cite{li2023blip} against its numerous medical-domain counterparts, which include models specialized for radiology (MedCLIP~\cite{wang2022medclip}, PubMedCLIP~\cite{eslami2023pubmedclip}), pathology (PLIP~\cite{huang2023visual}), and broad biomedical data (BioMedCLIP~\cite{zhang2023biomedclip}). Second, we assess the SigLIP family, represented by the generalist SigLIP~\cite{zhai2023sigmoid} and its medical adaptation, MedSigLIP~\cite{sellergren2025medgemma}. 
By pairing generalist and specialist models within each family, we can directly investigate the impact of domain-specific contrastive pretraining strategies.

\vspace{2mm}
\noindent\textbf{Generative VLMs}. Generative VLMs integrate image and text through multimodal transformers capable of producing free-form, human-like responses. For our analysis, we selected models from several prominent families, each represented by a generalist version and its specialist medical counterparts. These include the LLaVA family, comprising the generalist LLaVA-1.5~\cite{liu2023visual} and the medical pre-trained LLaVA-Med~\cite{li2023llava}; the Gemma family, including the generalist Gemma-3~\cite{team2025gemma} and specialist MedGemma~\cite{sellergren2025medgemma}; and the Qwen family, including the Qwen-VL~\cite{bai2025qwen2} and the specialist counterpart Lingshu~\cite{xu2025lingshu}.

\vspace{2mm}
\noindent\textbf{Lightweight adaptation}. We used task- and architecture-specific lightweight adaptation protocols for different models. For contrastive VLMs in disease diagnosis, the primary adaptation strategy was linear probing (LP): image embeddings were extracted from the frozen contrastive VLM backbone, and only a shallow task-specific linear classifier was trained on top of these embeddings. This protocol leaves the pretrained vision-language representation unchanged and updates only the downstream shallow classifier. For generative VLMs in VQA, adaptation was implemented as supervised fine-tuning (SFT) with parameter-efficient updates. Specifically, the multimodal projection module mapping visual features into the language-model space was trained, and LoRA~\cite{hu2022lora} adapters were applied to the language decoder while the remaining backbone parameters were kept fixed. These protocols were applied consistently to generalist and specialist counterparts within each comparison, so that observed differences reflected the deployment pathway under evaluation rather than differences in adaptation capacity.


\subsection{Agentic VQA inference}
\label{sec:agentic}

To address RQ4, we evaluated whether training-free multi-agent inference changes the comparative performance of generalist and specialist VLM backbones. This analysis was conducted only for generative VLMs on closed-form VQA tasks. We considered two representative agentic reasoning frameworks: MDAgents~\citep{kim2024mdagents} and UCAgents~\citep{feng2025ucagents}. MDAgents organizes inference into collaborative expert-style stages, whereas UCAgents is designed to encourage visually grounded reasoning and reduce textual drift. The agentic frameworks were used as inference-time scaffolds rather than as additional training procedures. No model parameters were updated in the agentic setting. Each framework was applied to the same underlying VLM backbones evaluated under standard prompting, and outputs were constrained to the same closed-form answer space.

\subsection{Statistical analysis}
To compare specialist medical against generalist models, we fit linear mixed-effects models (LMMs)~\citep{oberg2007linear} using restricted maximum likelihood (REML). LMMs were chosen to account for the non-independence arising from repeated measurements across shared datasets and shared model backbones, as well as the hierarchical structure of the benchmark in which multiple models are nested within model families. In every LMM, fixed effects encoded the experimental contrasts of scientific interest, while random intercepts for dataset, model family, and model nested within family absorbed per-dataset difficulty, family-level baseline differences, and residual model-specific bias, respectively.

\begin{figure}[t]
    \centering
    \includegraphics[width=\linewidth]{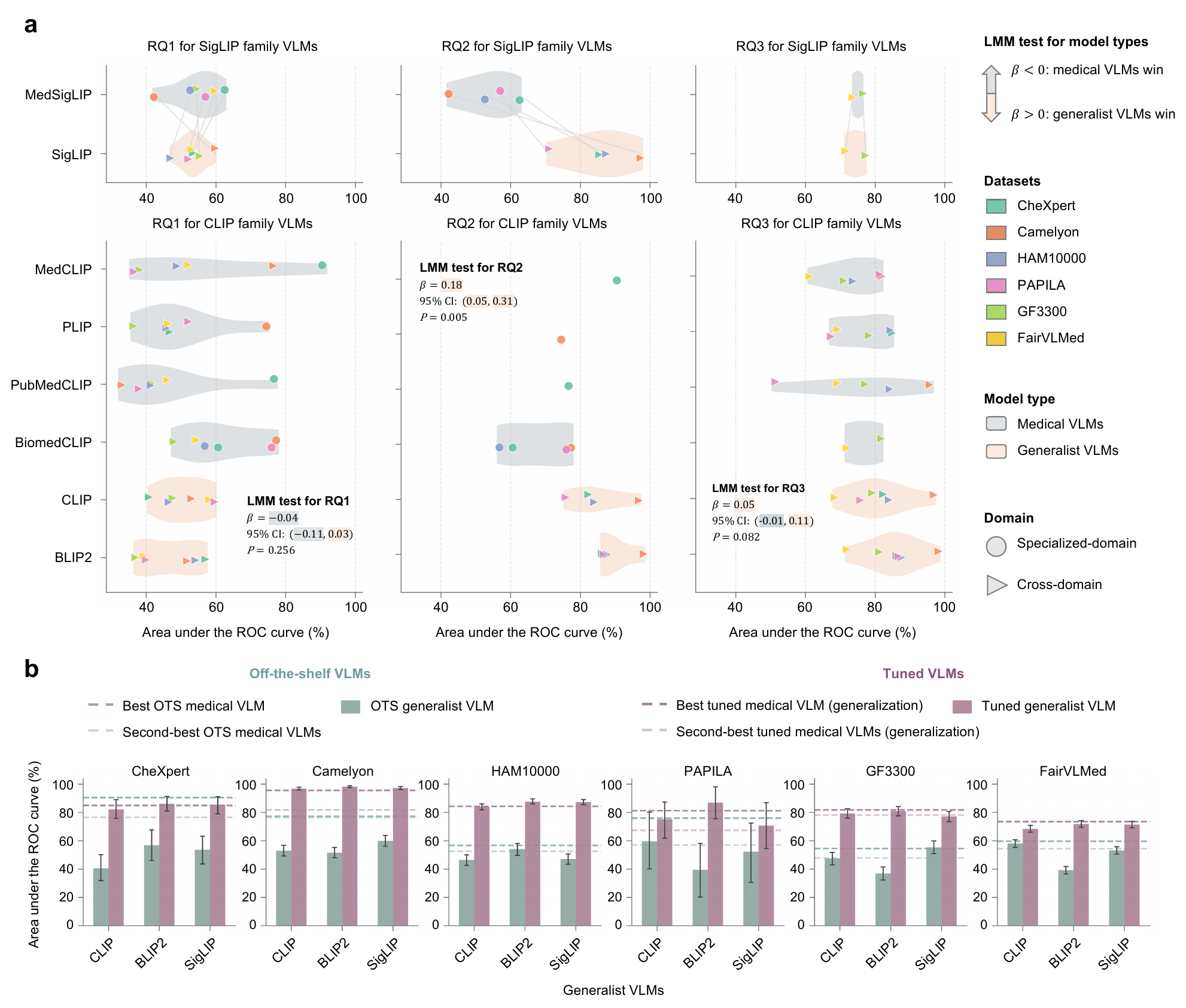}
    \caption{\textbf{Diagnosis results of contrastive VLMs.} \textbf{a.} Family-level comparisons across SigLIP- and CLIP-family VLMs for RQ1--RQ3. Each point denotes the mean model performance on a single dataset. The effect of model type (generalist versus medical) was estimated using linear mixed-effects models. The coefficient \(\beta\) is reported with 95\% confidence intervals and two-sided \(P\) values. \textbf{b.} Global cross-family comparisons across all contrastive VLMs. Generalist VLMs were compared with the best- and second-best-performing medical VLMs for RQ1--RQ3.}
    \label{fig:diagnosis}
\end{figure}

For each of RQ1–RQ3 (Fig.~\ref{fig:diagnosis} and Fig.~\ref{fig:vqa}), observations were pooled across model families, and a single LMM was fit per research question with model type (medical vs. general, medical as reference) as the sole fixed effect. The estimated coefficient thus quantifies the average performance gap between general and medical models after adjusting for the three random-intercept sources of variation. 
For RQ4 (Fig.~\ref{fig:vqa_agents}), we fit one LMM per agent, each restricted to zero-shot and that agent's observations, with fixed effects for model type, the usage of agentic systems (zero-shot as reference), and their interaction. The interaction coefficient is the effect of primary interest and quantifies the extent to which the agent disproportionately benefits generalist (relative to specialist) models. 
For every fitted model, we report the fixed-effect coefficient of interest $\beta$, its 95\% Wald confidence interval, and the associated two-sided $P$-value. All results are computed using non-parametric bootstrapping with 1,000 replicates.

\section{Results}
\label{exp}

\subsection{Diagnosis}
Fig.~\ref{fig:diagnosis} presents family-level and global comparisons of contrastive models for RQ1--RQ3 across disease diagnosis tasks on six datasets. 
The complete results are presented in Fig. S1 of the online supplementary material.
The key observations are summarized below.

\vspace{2mm}
\noindent\textbf{OTS specialist advantages are domain-specific (RQ1)}. 
In OTS medical applications, specialist medical VLMs tended to outperform generalist VLMs on average, but this advantage was neither robust nor statistically significant ($\beta=-0.04$, 95\% CI: $[-0.11, 0.03]$, $P=0.256$). The specialist advantage appeared more pronounced in the CLIP family than in the SigLIP family, where specialist and generalist distributions largely overlapped. 
Even within the CLIP family, however, the benefit was neither uniform nor consistently observed across datasets and modalities: most specialist models surpassed generalist counterparts only within subsets of their intended modality domains and underperformed on others. In the cross-family global comparison, the best-performing OTS medical VLM on each dataset exceeded the generalist baselines, yet on several datasets, including HAM10000 and FairVLMed, only the single strongest medical VLM did so.
These results indicate that the OTS benefit of specialist medical pretraining may be domain-dependent.


\vspace{2mm}
\noindent\textbf{Efficiently adapted generalists can offset OTS specialists (RQ2).}
We next assessed whether lightweight adaptation of generalist VLMs could provide a practical alternative to OTS specialist models. 
At the family level, adapted generalist VLMs significantly outperformed OTS specialist baselines ($\beta=0.18$, 95\% CI: $[0.05, 0.31]$, $P=0.005$). This overall pattern was visible within both CLIP and SigLIP families.
The global comparison showed a similar pattern. On CheXpert, the mean AUROC of fine-tuned generalist VLMs increased from 50.4\% to 84.5\%, approaching the best medical VLM. On PAPILA, the mean AUROC of fine-tuned generalists was 77.9\% compared to 75.9\% for the strongest medical VLM, indicating only a modest advantage. Across the remaining datasets, fine-tuned generalist VLMs substantially outperformed the strongest medical VLM. These findings support the practical value of lightweight local adaptation when modest task-specific data or compute are available.

\vspace{2mm}
\noindent\textbf{Cross-domain diagnosis does not consistently favour specialists (RQ3)}.
We further evaluated the cross-domain generalization of adapted generalist and specialist medical VLMs on unseen or clinically rare imaging modalities.
Pooled across both model families, adapted generalist models exhibited a modest but consistent advantage over specialist medical models trained under the same adaptation protocol ($\beta=0.05$, 95\% CI: $[-0.01, 0.11]$, $P=0.082$).
In the global cross-family comparison, all adapted generalist models approached or exceeded the best adapted specialist medical model across datasets and, in most cases, substantially outperformed the second-best medical model.
On PAPILA, for example, the three adapted generalist models achieved a mean AUROC of 77.9\%, with BLIP2 reaching 86.7\%, exceeding the best adapted medical model at 81.3\% and markedly surpassing the second-best medical model at 67.3\%. These findings suggest that large-scale medical-domain pretraining does not necessarily confer robust cross-domain diagnostic generalization and may, in some settings, be associated with reduced transferability to unseen medical imaging domains.


\begin{figure}[t]
    \centering
    \includegraphics[width=\linewidth]{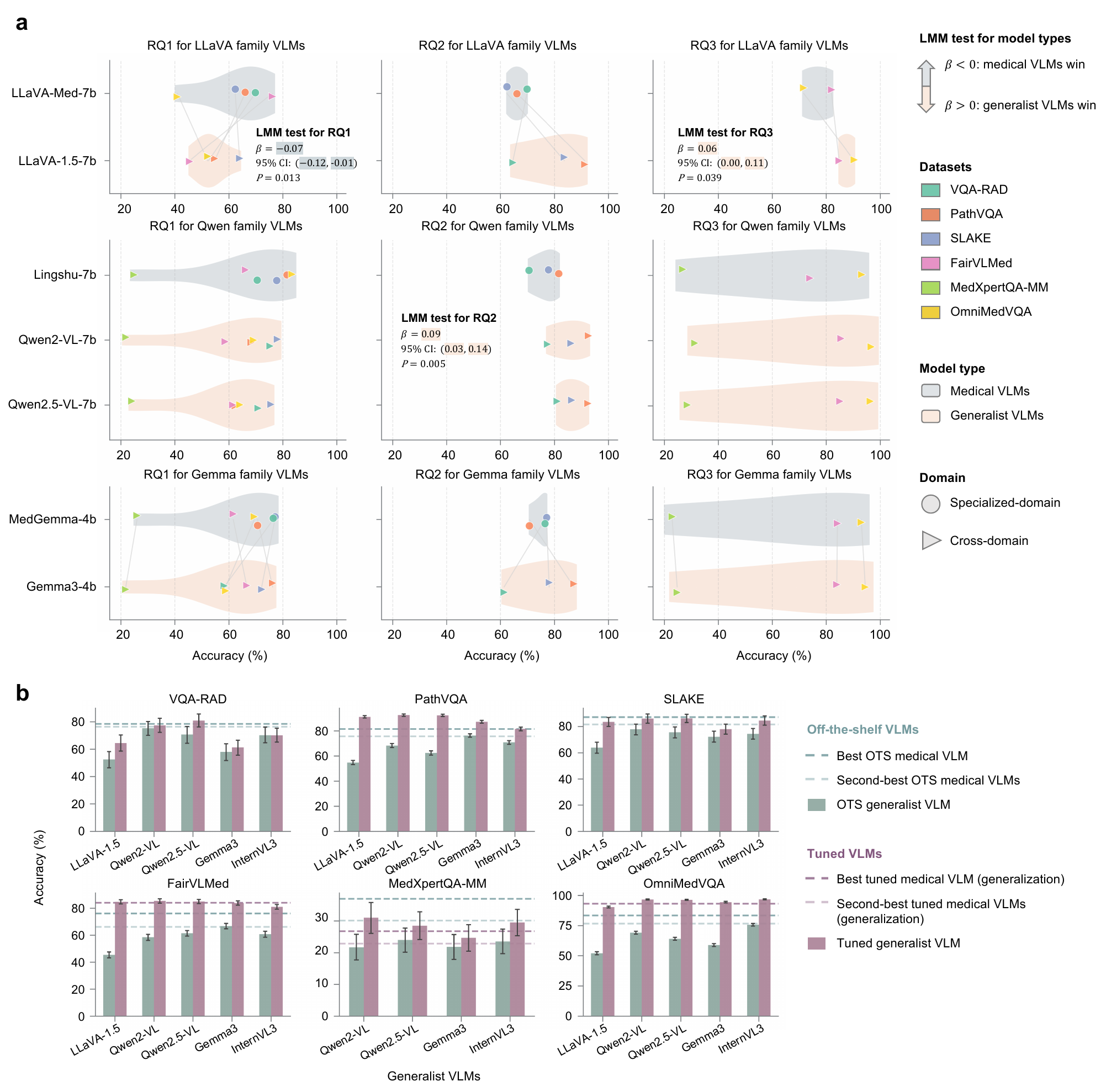}
    \caption{\textbf{VQA results of generative VLMs.} \textbf{a.} Family-level comparisons across LLaVA-, Qwen- and Gemma-family VLMs for RQ1--RQ3. Each point denotes the mean model performance on a single dataset. The effect of model type (generalist versus medical) was estimated using linear mixed-effects models. The coefficient \(\beta\) is reported with 95\% confidence intervals and two-sided \(P\) values. \textbf{b.} Global cross-family comparisons across all generative VLMs. Generalist VLMs were compared with the best- and second-best-performing medical VLMs for RQ1--RQ3. Frontier reasoning MLLMs, including o3 and Gemini 2.5 Pro, were included as strong generalist baselines to assess the potential of efficient adaptation.}
    \label{fig:vqa}
\end{figure}

\subsection{VQA}

Fig.~\ref{fig:vqa} presents family-level and global comparisons of generative models for RQ1--RQ3 across VQA tasks on six datasets. 
The detailed numeric results are presented in Fig. S2 in the online supplementary material.
Overall, the VQA results followed the same broad pattern as the diagnosis results. However, VQA differed in several respects.

\vspace{2mm}
\noindent\textbf{OTS specialist advantages are more pronounced in VQA (RQ1)}.
In the OTS setting, specialist medical VLMs showed a more stable and statistically significant advantage over generalist VLMs in VQA than in diagnosis ($\beta=-0.07$, 95\% CI: $[-0.12, -0.01]$, $P=0.013$).
This specialist advantage was more pronounced in the LLaVA family and Qwen. 
Unlike in diagnosis, the OTS specialist benefit remained apparent even on rare or unseen imaging modalities: for example, LLaVA-Med demonstrated strong generalization on FairVLMed, an unseen SLO fundus dataset. These findings suggest that large-scale medical pretraining may provide a clearer OTS benefit for MLLMs on generative VQA tasks than for contrastive diagnosis.

\vspace{2mm}
\noindent\textbf{Adapted generalists are competitive but less uniformly dominant in VQA (RQ2)}. 
We then examined whether lightweight adaptation could offset the OTS advantage of specialist medical MLLMs in VQA. 
Across model families, adapted generalist VLMs significantly outperformed OTS specialist baselines ($\beta=0.09$, 95\% CI: $[0.03, 0.14]$, $P=0.005$), although the gap was more nuanced than in the diagnosis setting. 
Within-family inspection suggested that this overall advantage was driven mainly by the LLaVA and Qwen families, whereas in the Gemma family the adapted generalist and OTS specialist distributions were largely overlapping. 
The global comparison presented the same mixed pattern: adapted generalist models approached or exceeded the best specialist medical model on most datasets. However, on MedXpertQA-MM, a heterogeneous and challenging benchmark, generalist models adapted with limited data remained below the strongest specialist model, the reasoning MLLM o3.

\vspace{2mm}
\noindent\textbf{Specialists remain less consistently robust in cross-domain VQA (RQ3)}. 
For cross-domain generalization, adapted generalist models exhibited more robust transfer performance than specialist medical models trained under the same adaptation protocol ($\beta=0.06$, 95\% CI: $[0.00, 0.11]$, $P=0.039$), consistent with the pattern observed in diagnosis tasks.
This pattern was observed across all three model families, with the clearest separation in the LLaVA and Qwen families. 
In the global cross-family comparison, most adapted generalist models outperformed the best cross-domain-adapted specialist medical model on each dataset.
These observations suggest that, as in diagnosis, medical-domain pretraining does not necessarily improve cross-domain transfer in MLLM-based VQA and may be associated with reduced generalization to unseen or heterogeneous medical imaging domains.

\begin{figure}[t]
    \centering
    \includegraphics[width=\linewidth]{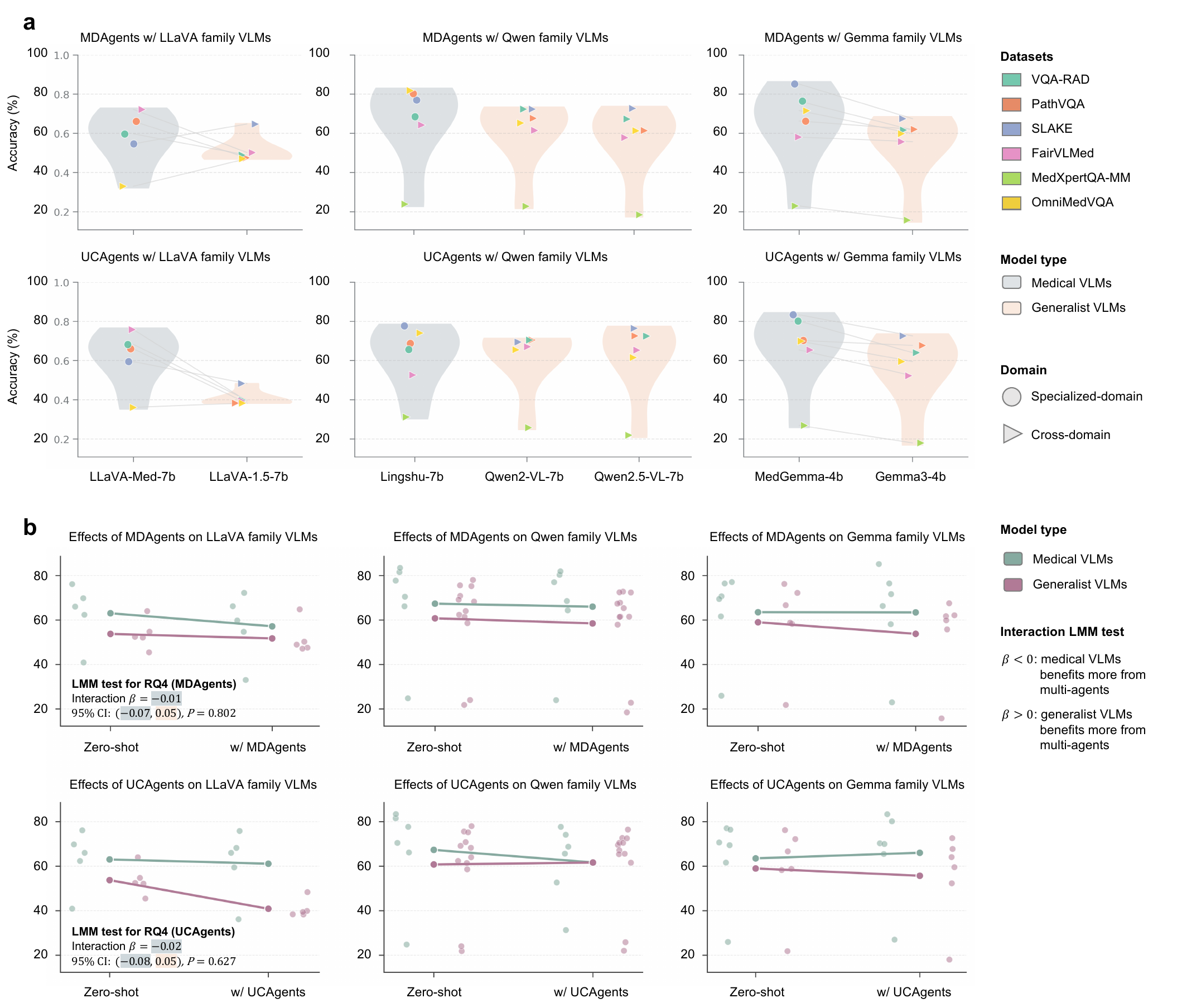}
    \caption{\textbf{Agentic VQA performance of generative VLMs integrated with multi-agent frameworks}. \textbf{a.} Family-level comparisons across LLaVA-, Qwen- and Gemma-family VLMs integrated with multi-agent frameworks. \textbf{b.} Performance gains after applying multi-agent frameworks across different model families. The interaction effect between model type (generalist versus medical) and the use of multi-agent frameworks was estimated using linear mixed-effects models. The coefficient \(\beta\) is reported with 95\% confidence intervals and two-sided \(P\) values.}
    \label{fig:vqa_agents}
\end{figure}

\subsection{Agentic VQA}

\noindent\textbf{Agentic reasoning yields limited, model-dependent gains and does not significantly differentiate generalist and specialist models (RQ4)}.
Fig. 4 shows the VQA performance of three generative model families integrated with two multi-agent frameworks that use multi-turn conversation and reasoning. The full results are presented in Fig. S3 of the online supplementary material. Across frameworks, the interaction between model type and the use of multi-agent frameworks was not statistically significant for either MDAgents ($\beta=-0.01$, 95\% CI: $[-0.07, 0.05]$, $P = 0.802$) or UCAgents ($\beta= -0.02$, 95\% CI: $[-0.08, 0.05]$, $P = 0.627$), indicating that specialist and generalist models did not differ significantly in the benefit they derived from agentic scaffolding.  
Overall gains were limited and heterogeneous. This pattern may reflect the fact that the current specialist medical VLMs are relatively small in scale (from several billion to a few tens of billions of parameters) and were not optimized for multi-step reasoning. Models of this size, or those lacking reasoning-oriented pretraining, generally struggled to benefit from agentic frameworks; in some cases, they even suffered performance degradation, exhibiting reasoning hallucinations and loss of information during long-context multi-turn dialogue. 

This heterogeneity was evident across specific agent–family combinations, with compatibility varying substantially across model families and agentic designs. For the earlier non-reasoning LLaVA-family MLLMs, multi-agent interaction introduced failure modes such as reasoning hallucinations and loss of long-context conversational memory, leading to negative performance gains for both generalist and medical models. In the Qwen family, generalist models benefited more than medical models when paired with UCAgents, whereas this effect was not visible with MDAgents. The Gemma family showed the opposite pattern, with medical models achieving slightly larger gains than their generalist counterparts under either framework. Together, these findings indicate that specialist and generalist models do not differ significantly under agentic frameworks, and that agentic VQA performance is instead shaped by the interaction between the agentic framework, model scale, architecture and pretraining regime—particularly pretraining for reasoning and chain-of-thought behaviour. This suggests that, for clinical tasks, building agentic systems on large-scale, general-purpose, reasoning-optimized generalist models may be a more suitable strategy.

\section{Discussion}
\label{discussion}

This study evaluated when available specialist medical VLMs offer practical advantages over generalist VLMs across four deployment-relevant settings: off-the-shelf use, lightweight adaptation, cross-domain transfer and training-free agentic reasoning. First, specialist advantages were domain-specific and inconsistent across families. In modality-matched settings, specialist models often outperformed generalist counterparts, consistent with reports that medical pretraining can improve radiology, pathology and other biomedical imaging performance~\citep{zhang2023biomedclip,li2023llava,sellergren2025medgemma}. However, these gains were not uniform across datasets or model families, reinforcing the need for task-specific evaluation rather than assumptions of clinical readiness based on pretraining domain alone~\citep{hager2024evaluation,wang2025perspective}.



Second, lightweight adaptation often changed the practical comparison. Adapted generalist VLMs frequently matched or exceeded off-the-shelf specialist baselines on specialised-domain tasks, indicating that modest local adaptation can offset much of the apparent specialist advantage when labelled data, validation capacity and fine-tuning infrastructure are available. This aligns with the view that foundation models should be evaluated in their intended deployment context, including local validation and updating~\citep{youssef2023external,feng2022clinical}. For resource-limited health systems, adapting a strong generalist VLM may therefore be a credible route for constrained imaging tasks.


Third, in the vast majority of settings, adapted generalist VLMs consistently achieved performance that was competitive with, or superior to, adapted specialist VLMs. This finding highlights the limitations of current specialist medical VLMs when generalizing to rare clinical modalities, and supports the use of non-specialized generalist models as a viable alternative for applications involving cross-modality generalization. Whereas many medical VLM benchmarks emphasize off-the-shelf performance, broad leaderboards or comparisons among specialist models~\citep{xu2025lingshu,royer2024multimedeval,jeong2024medical}, our cross-domain results test whether specialist pretraining improves domain depth at the expense of transfer breadth.


Fourth, under state-of-the-art multi-agent systems, generalist and specialist VLMs showed no significant difference in the benefit they derived, and overall gains were limited and inconsistent across model families.
The residual heterogeneity across families may further reflect differences in model scale, intrinsic reasoning ability, multi-turn dialogue capacity, architecture and pretraining strategy, highlighting the need for larger, reasoning-optimized medical backbones and dedicated clinical reasoning tools and agentic systems tailored to them.



These findings have implications for research, practice and policy. Benchmark design should distinguish off-the-shelf performance, lightweight adaptation, cross-domain transfer and possible pretraining exposure, because single leaderboard scores can obscure the sources of improvement. Future model reports should document pretraining scope, known benchmark exposure and matched or near-matched generalist comparators. Clinically, specialist medical VLMs may be prioritized for off-the-shelf use in modality-matched settings, whereas adapted generalist VLMs should be considered when labelled adaptation data and local validation are available. For procurement and regulation, a specialist label alone should not be treated as evidence of clinical suitability; adoption decisions should require transparent reporting of intended use, validation data, adaptation needs, dataset exposure and performance under clinically relevant conditions. 
Moreover, the benefit of agentic frameworks is governed less by domain specialization than by model scale and reasoning-oriented pretraining.  This points to a practical implication: for clinical tasks, constructing agents on large-scale (hundreds-of-billions to trillion-parameter), general-purpose models trained explicitly for reasoning and chain-of-thought behaviour may be a more promising strategy than relying on smaller specialist medical models. Dedicated evaluation of such backbones is an important direction for future work.

This study has limitations. 
Not all within-family comparisons were strictly matched for contrastive models; for instance, MedSigLIP is larger than SigLIP. The specialized models included here may have undergone minor changes in architecture and parameter count during their medical adaptation. We consider such modifications an integral part of specialization, and if anything, they strengthen our conclusions by placing the generalist models at a minor disadvantage in the comparison.
Pretraining exposure cannot be fully excluded for large public checkpoints, especially when training corpora are only partially disclosed. The specialised-domain and cross-domain taxonomy operates at the modality level and does not capture finer distribution shifts related to scanner, protocol, institution, population, disease prevalence or annotation. The benchmark also uses fixed adaptation sets and does not estimate the labelled-data threshold at which a generalist model matches or exceeds a specialist baseline. Finally, we focus on AUROC and closed-form VQA accuracy, without evaluating calibration, operating-point sensitivity and specificity, open-ended report generation, clinician interaction, prospective workflow performance, subgroup fairness or patient outcomes. Future work should address these gaps through stricter matched-pair comparisons, learning-curve analyses, expanded metrics including calibration and subgroup robustness, and prospective human-in-the-loop evaluation.

\section{Conclusion}
\label{conclusion}

In this paper, we introduced MedVLMBench and used it to ask when available specialist medical VLMs offer practical advantages over generalist VLMs.  Across 18 models and 12 public dataset components spanning off-the-shelf use, lightweight adaptation, cross-domain transfer and training-free agentic reasoning, specialist advantages were conditional rather than universal: they were most apparent for off-the-shelf, modality-matched deployment, but were largely offset once generalist models were lightly adapted, did not consistently extend to unfamiliar modalities, and did not translate into greater benefit under agentic frameworks.  These results caution against treating specialist medical pretraining as sufficient evidence of clinical suitability.  Model selection should instead be guided by the intended deployment pathway—including local data availability, adaptation and validation capacity, and performance under clinically relevant operating conditions—with strong generalist VLMs representing a credible and potentially more scalable alternative in cross-domain and resource-constrained settings.  Looking ahead, as clinical AI increasingly moves towards agentic applications, the limited and heterogeneous gains observed here suggest that realizing their potential will depend less on domain-specialized backbones than on larger, reasoning-oriented models and on agentic frameworks purpose-built for medical decision-making.

\section*{Data availability}

Data are available in public, open access repositories or may be obtained from third parties where access restrictions apply. \ours{} is built from previously released datasets.
Dataset sources and access links are provided in the appendix and should be accessed under the licences and conditions specified by the original data providers. 
The benchmark split files, preprocessing manifests and model prediction files will be made available in a public repository upon publication where permitted by dataset licences and access requirements. For datasets that cannot be redistributed, the release will provide deterministic split identifiers and preprocessing scripts so licensed users can reconstruct the evaluated benchmark.

\section*{Code availability}

Code for benchmark construction, model evaluation, prompting, lightweight adaptation and statistical analysis has been released at https://github.com/ubc-tea/MedVLMBench.

\section{Funding}
This work was supported by the Natural Sciences and Engineering Research Council of Canada (NSERC; Grant No. RGPIN-2022-0531) and the Research Grants Council of the Hong Kong Special Administrative Region, China (Project No. T45-401/22-N).


\clearpage

\pagenumbering{roman}
\section*{Supplemental Material}
\definecolor{lightgray}{gray}{0.95}

\setcounter{table}{0}
\setcounter{figure}{0}

\renewcommand{\thetable}{S\arabic{table}}
\renewcommand{\thefigure}{S\arabic{figure}}

\begin{center}
  \centering
  \captionof{table}{Contrastive VLMs for diagnosis in the paper, categorized by model family.}
  \label{tab:vlm_family_categorization}
  \footnotesize

  \setlength{\tabcolsep}{5pt} 
  \renewcommand{\arraystretch}{1.2} 
  \newcolumntype{L}[1]{>{\raggedright\arraybackslash}p{#1}}
  \newcolumntype{C}{>{\centering\arraybackslash}p{1.2cm}}
  \newcolumntype{Y}{>{\raggedright\arraybackslash}X} 

  \begin{tabularx}{\linewidth}{@{} L{2.4cm} C L{2.6cm} Y C C @{}}
    \toprule
    \textbf{VLM} & \textbf{Domain} & \textbf{Training modality} & \textbf{Training dataset(s)} & \textbf{Public} & \textbf{Model size} \\
    \midrule

    \multicolumn{6}{@{}l}{\textbf{CLIP family}} \\

    \texttt{CLIP}~\citesupp{radford2021learning} &
      Generalist &
      \textsc{Image--Text} (web) &
      400\,M WebImageText pairs &
      \raisebox{-0.2ex}{\cmark} &
      151.28M  \\

    \texttt{BLIP-2}~\citesupp{li2023blip} &
      Generalist &
      \textsc{Image--Text} (web) &
      COCO~\citesupp{lin2014microsoft}, Visual Genome~\citesupp{krishna2017visual},
      CC3M~\citesupp{sharma2018conceptual}, CC12M~\citesupp{changpinyo2021conceptual},
      SBU~\citesupp{ordonez2011im2text}, LAION-400M~\citesupp{schuhmann2021laion} (129\,M pairs) &
      \raisebox{-0.2ex}{\cmark} &
      1172.03M  \\

    \texttt{MedCLIP}~\citesupp{wang2022medclip} &
      Specialist Medical &
      Chest X-ray pairs &
      CheXpert~\citesupp{irvin2019chexpert}, MIMIC-CXR~\citesupp{johnson2019mimic} (570\,k pairs) &
      \raisebox{-0.2ex}{\cmark} &
      136.62M  \\

    \texttt{PubMedCLIP}~\citesupp{eslami2023pubmedclip} &
      Specialist Medical &
      Radiology images &
      ROCO (80\,k images from PubMed Central) &
      \raisebox{-0.2ex}{\cmark} &
      151.28M  \\

    \texttt{PLIP}~\citesupp{huang2023visual} &
      Specialist Medical &
      Pathology images &
      OpenPath (21\,k Twitter pathology pairs) &
      \raisebox{-0.2ex}{\cmark} &
      151.28M  \\

    \texttt{BioMedCLIP}~\citesupp{zhang2023biomedclip} &
      Specialist Medical &
      Multi-modality medical figures &
      PMC-15M (15.3\,M figure--caption pairs) &
      \raisebox{-0.2ex}{\cmark} &
      195.90M  \\

    \midrule[0.8pt]

    \multicolumn{6}{@{}l}{\textbf{SigLIP family}} \\

    \texttt{SigLIP}~\citesupp{zhai2023sigmoid} &
      Generalist &
      \textsc{Image--Text} (web) &
      WebLI~\citesupp{chen2022pali} (10\,B pairs) &
      \raisebox{-0.2ex}{\cmark} &
      203.16M \\

    \texttt{MedSigLIP}~\citesupp{sellergren2025medgemma} &
      Specialist Medical &
      Mixed (general + medical) &
      SigLIP and Gemma-3 mixtures; Med-Gemini corpus &
      \raisebox{-0.2ex}{\cmark} &
      878.30M  \\

    \bottomrule
  \end{tabularx}
\end{center}

\clearpage

\begin{center}
\centering
\captionof{table}{Generative VLMs for VQA in the paper, categorized by model family.}
\label{tab:gen_vlm_summary_part2}
\footnotesize

\setlength{\tabcolsep}{5pt}
\renewcommand{\arraystretch}{1.2}

\newcolumntype{L}[1]{>{\raggedright\arraybackslash}p{#1}}
\newcolumntype{D}[1]{>{\centering\arraybackslash}p{#1}}
\newcolumntype{Y}{>{\raggedright\arraybackslash}X}

\begin{tabularx}{\linewidth}{
  @{}
  L{2.2cm}
  D{1.2cm}
  L{2.2cm}
  L{2.2cm}
  Y
  D{0.8cm}
  D{1.2cm}
  @{}
}
\toprule
\textbf{VLM}
&
\textbf{Domain}
&
\textbf{Vision encoder}
&
\textbf{LLM backbone}
&
\textbf{Training dataset(s)}
&
\textbf{Public}
&
\textbf{Model size}
\\
\midrule

\multicolumn{7}{@{}l}{\textbf{\texttt{LLaVA} family}} \\

\texttt{LLaVA-1.5-7B}~~\citesupp{liu2023visual}
&
Generalist
&
CLIP ViT-L/14-336
&
Vicuna-v1.5 7B
&
Image--text feature alignment followed by multimodal
instruction tuning on general-domain visual instruction data
&
\cmark
&
7B
\\

\texttt{LLaVA-Med-7B}~~\citesupp{li2023llava}
&
Specialist Medical
&
CLIP ViT-L/14-336
&
Vicuna 7B
&
Biomedical figure--caption alignment followed by
biomedical visual instruction tuning using data derived
from PubMed Central
&
\cmark
&
7B
\\

\midrule[0.8pt]

\multicolumn{7}{@{}l}{\textbf{\texttt{Qwen} family}} \\

\texttt{Qwen2-VL-7B}~~\citesupp{wang2024qwen2}
&
Generalist
&
Dynamic-resolution ViT
&
Qwen2 7B
&
Large-scale image--text and video--text pre-training,
followed by multimodal supervised instruction tuning
&
\cmark
&
8B
\\

\texttt{Qwen2.5-VL-7B}~~\citesupp{bai2025qwen2}
&
Generalist
&
Dynamic-resolution ViT with window attention
&
Qwen2.5 7B
&
Large-scale multimodal pre-training followed by
supervised instruction tuning and preference optimization
&
\cmark
&
8B
\\

\texttt{Lingshu-7B}~~\citesupp{xu2025lingshu}
&
Specialist Medical
&
Qwen2.5-VL vision encoder
&
Qwen2.5 7B
&
Multi-stage medical alignment and instruction tuning
using medical image--text, interleaved multimodal,
and medical text data
&
\cmark
&
8B
\\

\midrule[0.8pt]

\multicolumn{7}{@{}l}{\textbf{\texttt{InternVL} family}} \\

\texttt{InternVL3-8B}~~\citesupp{zhu2025internvl3}
&
Generalist
&
InternViT-L (V2PE)
&
Qwen2.5 7B
&
Native joint multimodal pre-training, followed by
supervised fine-tuning and mixed-preference optimization
&
\cmark
&
7.94B
\\

\midrule[0.8pt]

\multicolumn{7}{@{}l}{\textbf{\texttt{Gemma} family}} \\

\texttt{Gemma-3-4B}~~\citesupp{team2025gemma}
&
Generalist
&
SigLIP ViT (400M, frozen)
&
Gemma 3 4B
&
Distillation using a mixture of multimodal and text data;
visual inputs represented using 256 vision tokens
&
\cmark
&
4.30B
\\

\texttt{MedGemma-4B}~~\citesupp{sellergren2025medgemma}
&
Specialist Medical
&
MedSigLIP ViT
&
Gemma 3 4B
&
Multi-stage training on medical multimodal and text
corpora; frozen vision encoder with 256 vision tokens
&
\cmark
&
4.97B
\\

\midrule[0.8pt]

\multicolumn{7}{@{}l}{\textbf{\texttt{OpenAI} family}} \\

\texttt{o3}~~\citesupp{openai2025o3o4mini}
&
Generalist
&
Proprietary multimodal encoder
&
OpenAI o3
&
Proprietary multimodal pre-training, reinforcement-learning
based alignment, and instruction tuning
&
\cmark
&
NA
\\

\midrule[0.8pt]

\multicolumn{7}{@{}l}{\textbf{\texttt{Gemini} family}} \\

\texttt{Gemini 2.5 Pro}~~\citesupp{comanici2025gemini}
&
Generalist
&
Proprietary multimodal encoder
&
Gemini 2.5 Pro
&
Joint multimodal pre-training followed by instruction
tuning and preference optimization
&
\cmark
&
NA
\\

\bottomrule
\end{tabularx}
\end{center}

 \clearpage





\begin{center}
  \centering
  \captionof{table}{Overview of the diagnosis datasets.}
  \label{tab:diagnosis-data}

  \small
  \setlength{\tabcolsep}{4pt}
  \renewcommand{\arraystretch}{1.2}

  \newcolumntype{U}{>{\footnotesize\raggedright\arraybackslash}X} 

  \begin{tabularx}{\linewidth}{@{}
      >{\bfseries}l
      l
      >{\raggedleft\arraybackslash}r
      >{\raggedleft\arraybackslash}r
      U @{}}
    \toprule
    Dataset & Modality & Train & Test & Download URL \\
    \midrule
    CheXpert~\citesupp{irvin2019chexpert}   & Radiology            & 223414 & 234  & \url{https://stanfordmlgroup.github.io/competitions/chexpert/} \\
    GF3300~\citesupp{luo2024harvard}     & Fundus photography   & 2706   & 594  & \url{https://ophai.hms.harvard.edu/datasets/harvard-gf3300/} \\
    PAPILA~\citesupp{kovalyk2022papila}     & Color retinal fundus & 364    & 56   & \url{https://www.nature.com/articles/s41597-022-01388-1} \\
    HAM10000~\citesupp{tschandl2018ham10000}   & Dermatology          & 8137   & 1811 & \url{https://dataverse.harvard.edu/dataset.xhtml?persistentId=doi:10.7910/DVN/DBW86T} \\
    FairVLMed~\citesupp{luo2024fairclip}  & SLO fundus           & 8254   & 1746 & \url{https://ophai.hms.harvard.edu/datasets/harvard-fairvlmed10k} \\
    Camelyon17~\citesupp{bandi2018detection} & Pathology            & 3680   & 920  & \url{https://camelyon17.grand-challenge.org/Data/} \\
    \bottomrule
  \end{tabularx}
\end{center}

\clearpage

\begin{center}
  \captionof{table}{Overview of the VQA datasets. Only yes/no and multiple-choice questions were included in the test splits.}
  \label{tab:vqa-data}

  \small
  \setlength{\tabcolsep}{4pt}
  \renewcommand{\arraystretch}{1.15}
  \newcolumntype{U}{>{\footnotesize\raggedright\arraybackslash}X}

  \begin{tabularx}{\linewidth}{@{}
      >{\bfseries}l
      l
      >{\raggedleft\arraybackslash}r
      >{\raggedleft\arraybackslash}r
      U@{}}
    \toprule
    Dataset & Modality & Train & Test & URL \\
    \midrule
    SLAKE~\citesupp{liu2021slake}         & CT, MRI, and X-ray          & 5980  & 416  & \url{https://huggingface.co/datasets/BoKelvin/SLAKE} \\
    PathVQA~\citesupp{he2020pathvqa}       & Pathology                   & 26373 & 3362  & \url{https://huggingface.co/datasets/flaviagiammarino/path-vqa} \\
    VQARAD~\citesupp{lau2018dataset}        & Radiology                   & 2244  & 251   & \url{https://huggingface.co/datasets/flaviagiammarino/vqa-rad} \\
    FairVLMed~\citesupp{luo2024fairclip}     & Fundus imaging (SLO)        & 17994 & 1996  & \url{https://ophai.hms.harvard.edu/datasets/harvard-fairvlmed10k} \\
    MedXpertQA-MM~\citesupp{zuo2025medxpertqa} & Multimodal (radiology, pathology, clinical) & 1600 & 400 & \url{https://huggingface.co/datasets/TsinghuaC3I/MedXpertQA} \\
    OmniMedVQA~\citesupp{hu2024omnimedvqa}    & Multimodal medical imaging  & 2677  & 6186  & \url{https://huggingface.co/datasets/foreverbeliever/OmniMedVQA} \\
    \bottomrule
  \end{tabularx}
\end{center}




\clearpage

\definecolor{medicalcolor}{HTML}{86AAA0}
\definecolor{generalistcolor}{HTML}{99648A}

\begin{figure}[t]
    \centering
    \includegraphics[width=\linewidth]{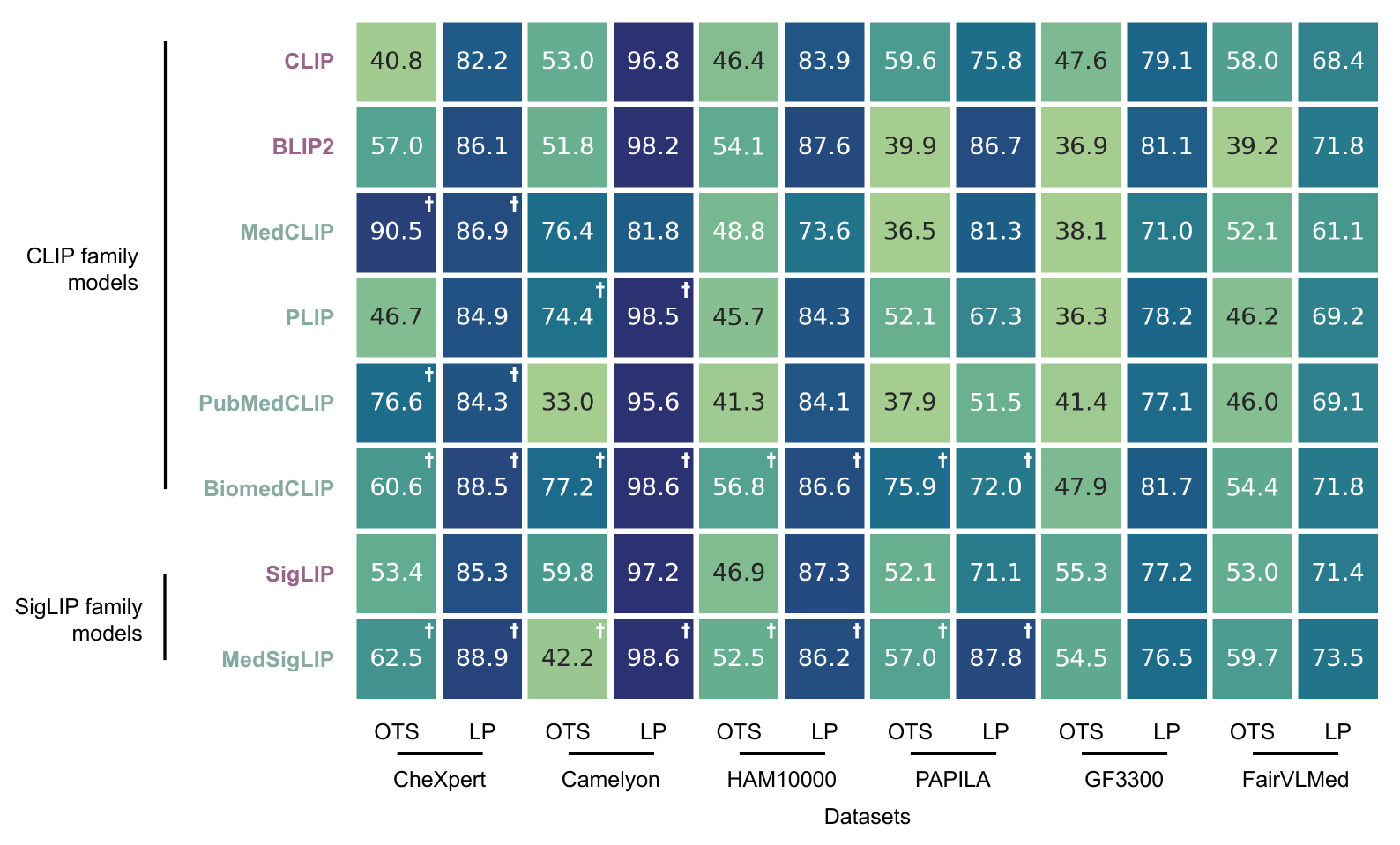}
    \caption{\textbf{Diagnosis results for contrastive VLMs.} Performance is reported as area under the receiver operating characteristic curve (AUROC, \%). Rows correspond to evaluated model checkpoints, grouped by model family and model type, and columns correspond to dataset-adaptation settings. \textcolor{medicalcolor}{Specialist medical} and \textcolor{generalistcolor}{generalist} models are indicated by different colors. \textit{OTS} denotes off-the-shelf evaluation without task-specific training. \textit{LP} denotes linear probing. Dagger symbols denote specialized-domain settings, in which the dataset modality was seen during the corresponding model's pretraining stage.}
    \label{fig:heatmap_diagnosis}
\end{figure}

\clearpage

\begin{figure}[t]
    \centering
    \includegraphics[width=\linewidth]{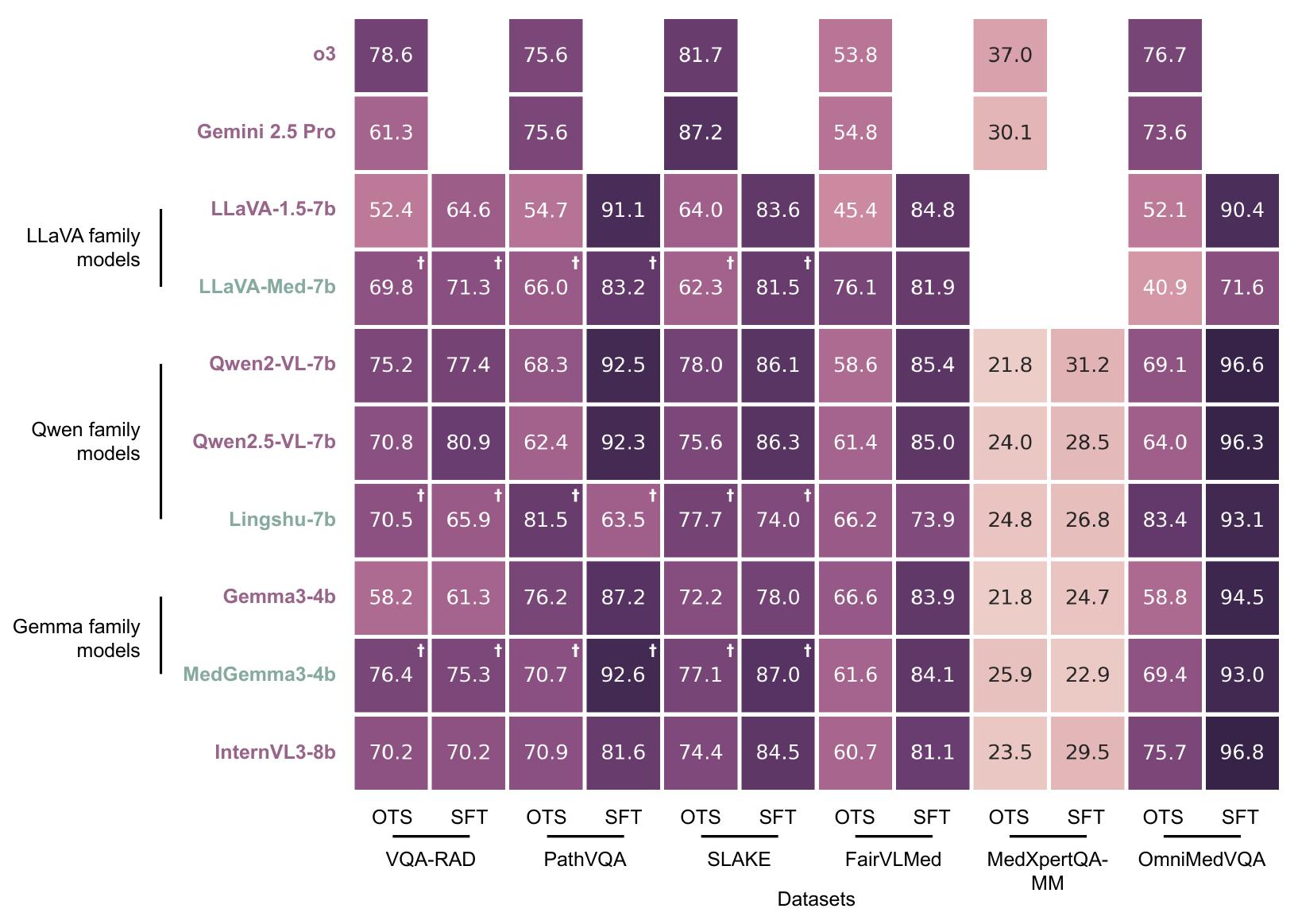}
    \caption{\textbf{VQA results for generative VLMs.} Performance is reported as accuracy (\%). Rows correspond to evaluated model checkpoints, grouped by model family and model type, and columns correspond to dataset-adaptation settings. \textcolor{medicalcolor}{Specialist medical} and \textcolor{generalistcolor}{generalist} models are indicated by different colors. \textit{OTS} denotes off-the-shelf evaluation without task-specific training. \textit{SFT} denotes supervised fine-tuning. Dagger symbols denote specialized-domain settings, in which the dataset modality was seen during the corresponding model's pretraining stage. LLaVA family models do not support multi-image inputs, which are required in the MedXpertQA-MM dataset.}
    \label{fig:heatmap_vqa}
\end{figure}

\clearpage

\begin{figure}[t]
    \centering
    \includegraphics[width=\linewidth]{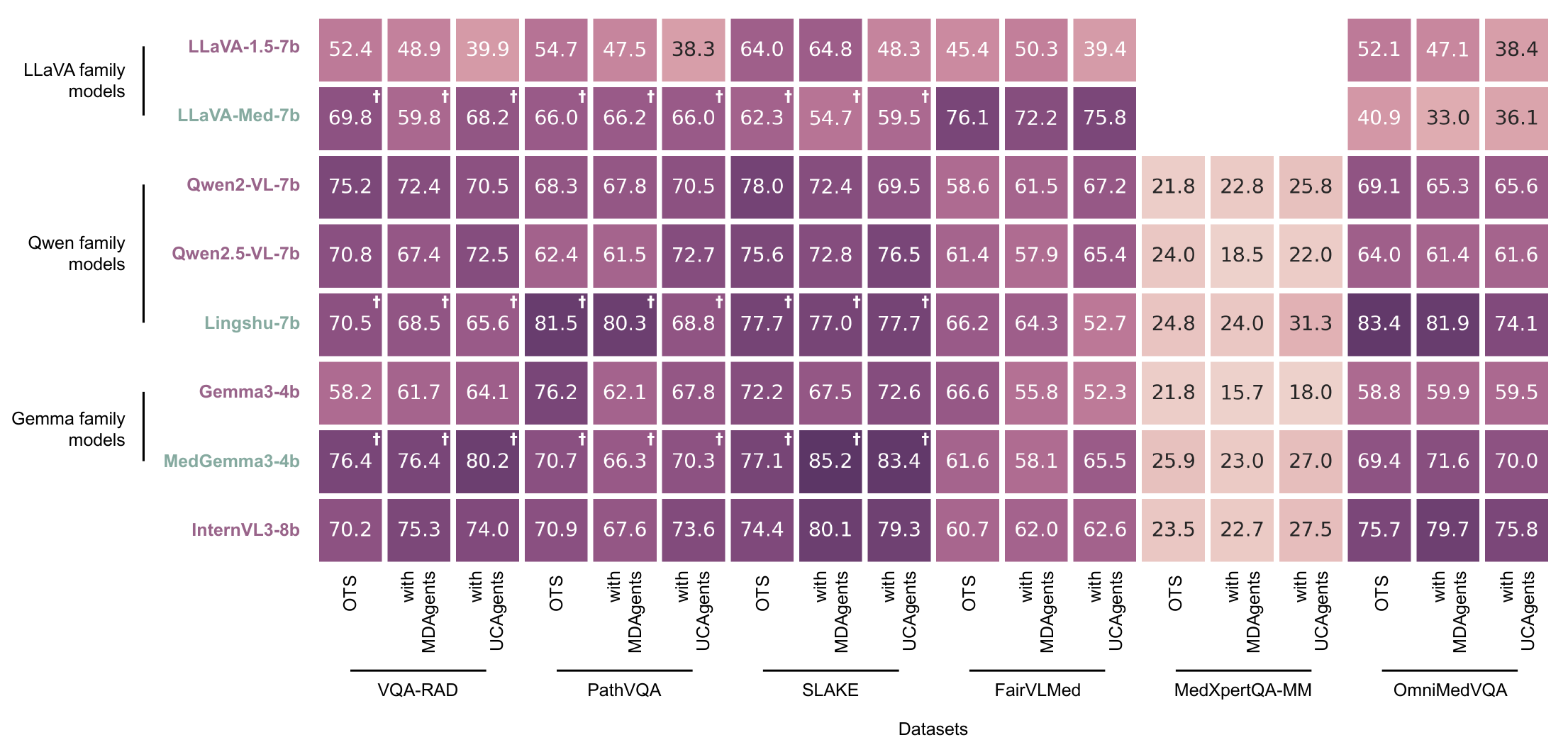}
    \caption{\textbf{Agentic VQA results}. Performance is reported as accuracy (\%). Rows correspond to evaluated model checkpoints, grouped by model family and model type, and columns correspond to different settings. \textcolor{medicalcolor}{Specialist medical} and \textcolor{generalistcolor}{generalist} models are indicated by different colors. \textit{OTS} denotes off-the-shelf evaluation without task-specific training. Two multi-agent frameworks, MDAgents and UCAgents, were included for agentic VQA evaluation. Dagger symbols denote specialized-domain settings, in which the dataset modality was seen during the corresponding model's pretraining stage.}
    \label{fig:heatmap_vqa_agents}
\end{figure}

\clearpage

\begin{figure}[H]
\centering
\begin{tcolorbox}[
    colframe=black!75!white,
    colback=blue!5!white,
    sharp corners=all,
    boxrule=0.6mm,
    rounded corners=all,
    width=\textwidth,
    title={\textbf{Prompt for VQA dataset construction}},
    fonttitle=\bfseries
]
\small\ttfamily
You are tasked with generating questions and answers for a Visual Question Answering (VQA) dataset based on medical notes associated with SLO fundus images. For each image paired with medical notes, you will create one open-ended question and one closed (yes/no) question. Follow these instructions carefully:

Use the provided medical notes as context to design both questions.

Ensure the open-ended question requires a descriptive answer based on the notes. The question and answer to it should be concise (within 20 words).

Ensure the closed question requires a simple yes or no answer, clearly derived from the notes.

It is important to remember that you are designing questions for VQA. The questions should be answerable from the SLO fundus image only, not from other information such as medical history or results of other tests that may appear in the notes. Remember that people answering the questions you provide do not have access to any other information except for the SLO fundus image.

Your response must strictly follow this format (no additional text):

Open Question: \textless Your open-ended question here\textgreater

Open Answer: \textless Your descriptive answer here\textgreater

Closed Question: \textless Your yes/no question here\textgreater

Closed Answer: \textless Yes or No\textgreater
\end{tcolorbox}

\caption{Prompt template used to generate candidate FairVLMed VQA question--answer pairs from paired SLO fundus images and clinical notes. The template constrained GPT-4 to generate one concise open-ended question--answer pair and one yes/no question--answer pair intended to be answerable from the image alone. Generated items were subsequently human validated before inclusion in the benchmark.}
\label{fig:vqa_prompt}
\end{figure}

\end{document}